\documentclass[preprint,12pt]{elsarticle}

\usepackage{amsthm}
\usepackage{amsmath,amssymb}

\renewcommand{\vec}[1]{\stackrel{\rightharpoonup}{#1}}

\journal{Annals of Pure and Applied Logic}

\begin{document}

\begin{frontmatter}



\title{A New Perspective for \\
Hoare's Logic and Peano's Arithmetic \tnoteref{t1}}

\tnotetext[t1]{Supported by the National Natural Science Foundation of China under
Grant No. 61272135, and National Science and Technology
Major Project of China under Grant No. 2012ZX01039-004.}


\author{Zhaowei Xu}
\ead{xuzw@ios.ac.cn}
\address{State Key Laboratory of Computer Science, Institute of Software,\\
Chinese Academy of Sciences, Beijing, China, 100190\\
University of Chinese Academy of Sciences\\}

\begin{abstract}
Hoare's logic is an axiomatic system of proving programs correct, which has been extended to be a separation logic to reason about mutable heap structure. We develop the most fundamental logical structure of strongest postcondition of Hoare's logic in Peano's arithmetic $PA$. Let $p\in L$ and $S$ be any while-program. The arithmetical definability of $\textbf{N}$-computable function $f_S^{\textbf{N}}$ leads to separate $S$ from $SP(p,S)$, which defines the strongest postcondition of $p$ and $S$ over $\textbf{N}$, achieving an equivalent but more meaningful form in $PA$. From the reduction of Hoare's logic to PA, together with well-defined underlying semantics, it follows that Hoare's logic is sound and complete relative to the theory of $PA$, which is different from the relative completeness in the sense of Cook. Finally, we discuss two ways to extend computability from the standard structure to nonstandard models of $PA$.
\end{abstract}


\end{frontmatter}


\section{Introduction}
Hoare's logic is a formal system for the derivation of statements about the partial correctness of programs \cite{apt_1}; it was introduced by Hoare \cite{hoare_1} and studied by Cook\cite{cook_1}, Lipton\cite{lipton_1}, Wand\cite{wand_1} and Clarke\cite{clarke_1}. Separation logic is a spatial logic for reasoning about mutable heap structure (\cite{reynolds_1,o'hearn_1,parkinson_1}), which is an extension to Hoare's logic to describe the applications of programs on the heap structure and the reasoning about memory update. It would be interesting to reconsider Hoare's logic and the computability induced by the accessability relations on models of Hoare's logic, since Hoare's logic, taken as a modal logic \cite{modal_1}, defines its model $\textbf{M}'$ on the assignments of the model $\textbf{M}$ of the first-order languages based on which Hoare's logic is defined, which leads to the special form of the completeness theorem different from that of the first-order logic.

Hoare's logic for the set $WP$ of all while-programs with first-order assertion language $L$ and first-order specification $T$ we denote $HL(T)$. The formulas in Hoare's system are triples of the form $\{p\}S\{q\}$, where $p,q\in L$ and $S\in WP$. We also call the triple $\{p\}S\{q\}$ a specified or asserted program. The specified program $\{p\}S\{q\}$ is true in a first-order model $\textbf{M}$, denoted by $\textbf{M}\models \{p\}S\{q\}$, if and only if for any initial assignment of the program variables over $\textbf{M}$, if $p$ holds  and $S$ is executed, then either $S$ will fail to terminate or $q$ will be satisfied by the final assignment defined by the program modality over $\textbf{M}$. Call the language $L$ expressive relative to model $\textbf{M}$ if for all $p\in L$ and $S\in WP$ there exists $q\in L$ which defines the strongest postcondition of $p$ and $S$ relative to $\textbf{M}$. If $L$ is expressive relative to $\textbf{M}$, then $\textbf{M}\models \{p\}S\{q\}$ implies $HL(Th(\textbf{M}))\vdash \{p\}S\{q\}$, where $p,q\in L$, $S\in WP$ and by $Th(\textbf{M})$ we mean the set of all true $L$-sentences in $\textbf{M}$. This is the so-called relative completeness(i.e. relative to some model) in the sense of Cook.

In the following, let $L$ be the language of arithmetic. Since $L$ is expressive relative to the standard structure $\textbf{N}$, then any true specified program can be proved in $HL(Th(\textbf{N}))$. Bergstra and Tucker\cite{bergstra_1} studied verification on an entirely proof-theoretic basis, where the facts about arithmetic must be formally deduced from Peano's arithmetic $PA$ or its refinement, and not popped from the oracle $Th(\textbf{N})$. The main theorem proved by them is listed as follows.

\newtheorem{bergstra_theorem}{Theorem}[section]
\begin{bergstra_theorem}[{Bergstra, et.al [1983]}]\label{bergstra_theorem}
Given an assertion $p\in L$ and program $S\in WP$ one can effectively calculate
an assertion $SP(p,S)\in L$ such that

(1) $SP(p,S)$ defines the strongest postcondition of S relative to p on the set of states
over $N$;

(2) $HL(PA)\vdash \{p\}S\{SP(p,S)\}$.
And, for any refinement T of Peano arithmetic, including PA itself,

(3) $HL(T)\vdash \{p\}S\{q\}$ if and only if $T\vdash SP(p,S)\rightarrow q$.
\end{bergstra_theorem}
The requirement of expressiveness restricts the interpretation of $PA$ to only the standard structure $\textbf{N}$ up to isomorphism \cite{lipton_1}. Therefore, relative to $\textbf{N}$, Bergstra and Tucker proved that $HL(PA)$ can be reduced to $PA$, which is their original intent. Closer scrutiny of their argumentation reveals that Theorem \ref{bergstra_theorem} also holds relative to the theory of $PA$(i.e. relative to all models of $PA$) under the assumption that they generalized the classical coding function, which is used to code a finite sequence of natural numbers, to code an $\textbf{M}$-finite sequence of numbers for any model $\textbf{M}$ of $PA$(i.e. generalized coding function), where by $\textbf{M}$-finite we mean the sequence has 'length' corresponding to elements in $\textbf{M}$. In the sequel, we develop our theoretical results relative to the theory of $PA$ under this assumption.

The semantical definition of strongest postcondition is well formalized in Cook\cite[p85]{cook_1}. Bergstra, et.al\cite[p271]{bergstra_1} gives the formal definition of strongest postcondition of $p\in L$ and $S\in WP$ relative to $\textbf{N}$, i.e. $SP(p,S)$ mentioned above, and proves some of its important properties as a theorem in $PA$. In this paper, we develop the most fundamental logical structure of strongest postcondition of Hoare's logic applied to arithmetic as it is defined by Peano's axioms, through which we can explain strongest postcondition more intuitively at the syntactic level. In principle, the $\textbf{N}$-computability induced by while-programs is equivalent to that of recursive functions, which are arithmetically definable \cite[p199]{c. and l.}, hence, we can arithmetically define $\textbf{N}$-computable function $f^{\textbf{N}}_S$ induced by while-program $S$ concretely. The arithmetical definability of $f^{\textbf{N}}_S$ leads to separate $S$ from $SP(p,S)$, achieving an equivalent but more meaningful form in $PA$, which is illustrated as follows.

\newtheorem{separation_theorem}[bergstra_theorem]{Theorem}
\begin{separation_theorem}[Separation Theorem]
For any $n$-place formula $p(\vec{x})\in L$, and $n$-place program $S\in WP$,
\begin{equation*}
    PA\vdash SP(p,S)(\vec{x})\leftrightarrow \exists \vec{u}\big(p(\vec{u}/\vec{x})\wedge \alpha_S(\vec{u}/\vec{x},\vec{x}/\vec{y})\big),
\end{equation*}
where $\alpha_S(\vec{x},\vec{y})$(resp. $SP(p,S)(\vec{x})$) arithmetically defines $f^{\textbf{N}}_S$(resp. $sp_{\textbf{N}}(p,S)$), $p(\vec{u}/\vec{x})$, i.e. $p(u_1/x_1,\ldots,u_n/x_n)$, stands for the result of substituting $u_i$ for each free occurrence of $x_i$ in $p$ for all $1\leq i\leq n$ and $\alpha_S(\vec{u}/\vec{x},\vec{x}/\vec{y})$ is defined analogously.

\end{separation_theorem}

The Separation Theorem captures the essence of the expressiveness condition. Replacing $SP(p,S)$ by $\exists \vec{u}\big(p(\vec{u}/\vec{x})\wedge \alpha_S(\vec{u}/\vec{x},\vec{x}/\vec{y})\big)$ in statement $(3)$ of Theorem \ref{bergstra_theorem}, one can not just obtain a conceptual clarity, but also simplify the strongest postcondition calculus inside considerably. Then it follows the following corollary.

\newtheorem{from_hl_to_pa}[bergstra_theorem]{Corollary}
\begin{from_hl_to_pa}\label{from_hl_to_pa_label}
For any $n$-place formulas $p(\vec{x}),q(\vec{x}) \in L$ and $n$-place program $S \in WP$,
\begin{displaymath}
HL(PA)\vdash \{p\}S\{q\} \mbox{ iff } PA \vdash p(\vec{x})\wedge \alpha_S(\vec{x},\vec{y})\rightarrow q(\vec{y}/\vec{x}),
\end{displaymath}
where $\alpha_S(\vec{x},\vec{y})$ arithmetically defines $f^{\textbf{N}}_S$, $q(\vec{y}/\vec{x})$, i.e. $q(y_1/x_1,\ldots,y_n/x_n)$, stands for the result of substituting $y_i$ for each free occurrence of $x_i$ in $q$ for all $1\leq i\leq n$.
\end{from_hl_to_pa}

The corollary reduces Hoare's logic grounded on the theory of $PA$ to the first-order logic $PA$. If we can extend computability induced by $S$ from the standard structure to nonstandard models of $PA$ with the same $\Sigma_1$-definability, that is $\textbf{M}$-computable function $f_S^{\textbf{M}}$ is defined by $\alpha_S(\vec{x},\vec{y})$ in nonstandard model $\textbf{M}$ of $PA$, then we can well define $\textbf{M}\models \{p\}S\{q\}$ for any model $\textbf{M}\models PA$, which we denote by $HL(PA) \models \{p\}S\{q\}$. With the well-defined underlying semantics, it easily deduces that $HL(PA) \models \{p\}S\{q\}$ iff $PA\models p(\vec{x})\wedge \alpha_S(\vec{x},\vec{y})\rightarrow q(\vec{y}/\vec{x})$. From G$\ddot{o}$del's Soundness and Completeness Theorem, together with Corollary \ref{from_hl_to_pa_label}, it follows that $HL(PA) \models \{p\}S\{q\}$ iff $HL(PA)\vdash \{p\}S\{q\}$. Consequently, Hoare's logic is sound and complete relative to the theory of $PA$, which is different from the relative completeness. More remarks about how to build the underlying semantics are left for the last section.

The paper is organized as follows: In Preliminaries, we give the formal description of Peano's arithmetic and coding functions. In section 3, we define Hoare's logic based on $PA$, its syntax and semantics. In section 4, we first give the definition of the $\textbf{N}$-computable function and argue its arithmetical definability; Then, we inductively construct $SP(p,S)$ like that in \cite[p273]{bergstra_1}; Finally, with the generalized coding function, we prove the Separation Theorem. In the last section, we discuss the extended computability and our future work.

\section{Preliminaries}

\subsection{Peano's arithmetic}
By $PA$ we denote the theory of Peano's arithmetic, which is the set of all sentences of the language of arithmetic that are provable from Peano's axioms. Also, in this paper, we treat $PA$ as the logical system of Peano's arithmetic. The desired data type semantics is the standard model $\textbf{N}$. The domain of $\textbf{N}$ is the set $\omega$ of natural numbers and its primitive operations are the \emph{addition} $x+y$ and \emph{multiplication} $x\cdot y$; $<$ is the linear ordering relation on $N$; $0, 1\in \omega$ are two distinguished elements. We shall use these notations for the function symbols, relation symbol and constant symbols in $L$. Let $L=\{+,\cdot,<,\textbf{0},\textbf{1}\}$. List the axioms of Peano's arithmetic as follows.
\begin{description}
  \item[(S1)] $x+\textbf{1}\neq \textbf{0}$;
  \item[(S2)] $x+\textbf{1}=y+\textbf{1}\rightarrow x=y$;
  \item[(S3)] $x+\textbf{0}=x$;
  \item[(S4)] $x+(y+\textbf{1})=(x+y)+\textbf{1}$;
  \item[(S5)] $x\cdot \textbf{0}=\textbf{0}$;
  \item[(S6)] $x\cdot (y+\textbf{1})=(x\cdot y)+x$;
  \item[(S7)] $\varphi(\textbf{0},\vec{y})\wedge \forall x\big(\varphi(x,\vec{y})\rightarrow \varphi(x+\textbf{1},\vec{y})\big)\rightarrow \forall x\varphi(x,\vec{y})$, where $\varphi(x,\vec{y})$ is a formula in $L$.
\end{description}

For simplicity, we abbreviate $\overbrace{\textbf{1}+\ldots+\textbf{1}}^{n\ \textbf{1}'s}$ as $\textbf{n}$. From the above Peano's axioms, we may observe that equations (S3)-(S6) alone define $\textbf{N}$ under initial algebra semantics and so we may consider (S1) and (S2) as additions, making a first refinement of the standard algebraic specification for arithmetic, desired to rule out finite models. The theoretical objective of adding the induction axiom (S7) is self-evident: one wants to generate all assertions which make statements about $\textbf{N}$ which can be based on its simple arithmetical operators and which can be proved by the principle of induction. For example, one can obtain facts about the ordering $x< y$ of natural numbers by using the formula $\exists z ( z\neq \textbf{0}\wedge x+z=y )$.

Let $\textbf{M}$ = ( $M$, $+^{\textbf{M}}$, $\cdot^{\textbf{M}}$, $<^{\textbf{M}}$, $\textbf{0}^{\textbf{M}}$, $\textbf{1}^{\textbf{M}}$ ) be a countable model of Peano arithmetic, i.e. $\textbf{M}\models \varphi$ for each axiom $\varphi$ of $PA$, where $M$ is the universe of $\textbf{M}$, and $I$ is an interpretation such that
\begin{eqnarray*}
  I(+,\textbf{M}) &=& +^{\textbf{M}}; \\
  I(\cdot,\textbf{M}) &=& \cdot^{\textbf{M}}; \\
  I(<,\textbf{M}) &=& <^{\textbf{M}}; \\
  I(\textbf{0},\textbf{M}) &=& \textbf{0}^{\textbf{M}}; \\
  I(\textbf{1},\textbf{M}) &=& \textbf{1}^{\textbf{M}}.
\end{eqnarray*}

If $\textbf{M}$ is not isomorphic to the standard model $\textbf{N}$ = ($N$, $+^{\textbf{N}}$, $\cdot^{\textbf{N}}$, $<^{\textbf{N}}$, $0$, $1$ ), then $\textbf{M}$ is nonstandard. There is a submodel $\textbf{N}^{\textbf{M}}$ = ( $N^{\textbf{M}}$, $+^{\textbf{N}^{\textbf{M}}}$, $\cdot^{\textbf{N}^{\textbf{M}}}$, $<^{\textbf{N}^{\textbf{M}}}$, $\textbf{0}^{\textbf{N}^{\textbf{M}}}$, $\textbf{1}^{\textbf{N}^{\textbf{M}}}$ ) of $\textbf{M}$ such that $\textbf{N}^{\textbf{M}}$ is isomorphic to $\textbf{N}$, where $N^{\textbf{M}}$ = \{$ I(\textbf{n},\textbf{M}):n\in N $\} is the universe of $\textbf{N}^{\textbf{M}}$ and
\begin{eqnarray*}
  +^{\textbf{N}^{\textbf{M}}} &=& +^{\textbf{M}}\upharpoonright N^{\textbf{M}}; \\
  \cdot^{\textbf{N}^{\textbf{M}}} &=& \cdot^{\textbf{M}}\upharpoonright N^{\textbf{M}}; \\
  <^{\textbf{N}^{\textbf{M}}} &=& <^{\textbf{M}}\upharpoonright N^{\textbf{M}}; \\
  \textbf{0}^{\textbf{N}^{\textbf{M}}} &=& \textbf{0}^{\textbf{M}}; \\
  \textbf{1}^{\textbf{N}^{\textbf{M}}} &=& \textbf{1}^{\textbf{M}}.
\end{eqnarray*}
We call elements in $N^{\textbf{M}}$ standard, and nonstandard otherwise. There exists a nonstandard model of $PA$, which is a direct consequence of the compactness theorem(see the argument of Corollary 25.3 in \cite[p306]{c. and l.}). For more information about nonstandard models, refer to Chapter 25 in \cite{c. and l.}.

In this paper, we need to understand the order in nonstandard models of $PA$. Let $K$ be the set consisting of all natural numbers together with all pairs $(q,a)$ where $q$ is a rational number and $a$ an integer. Let $<_K$ be the order on $K$ in which the natural numbers come first, in their usual order, and the pairs afterward, ordered as follows: $(q,a) <_K (r,b)$ if and only if $q<r$ in the usual order on rational numbers, or $q=r$ and $a<b$ in the usual order on integers. Then, the argument of Theorem 25.1 in \cite[p304]{c. and l.} implies that
\newtheorem{lemma_order}{Lemma}[subsection]
\begin{lemma_order}
  The order relation on any enumerable nonstandard model of $PA$ is isomorphic to the ordering $<_K$ of $K$.
\end{lemma_order}

The semantics of the first-order logic has its standard definition in model theory and we assume to be understood. The validity of $p\in L$ over model $\textbf{M}$ we write $\textbf{M}\models p$. Let $T$ be a theory of $L$. We write $T\models p$ to mean that for every $\textbf{M}\models T$, $\textbf{M}\models p$. From G\"{o}del's Soundness and Completeness Theorem,  it deduces that $T\vdash p$ if and only if $T\models p$.

\subsection{Coding functions}

A function $f:N^n\rightarrow N$ is \emph{arithmetically definable}, or \emph{arithmetical} for short, if and only if there is a formula $F(\vec{x},y)$ in $L$ such that for any $\vec{a}\in N^n, b\in N$ we have $f(\vec{a})=b$ if and only if $\textbf{N}\models F(\vec{\textbf{a}},\textbf{b})$. Boolos, et.al\cite[p204]{c. and l.} writes that:

\newtheorem{recursive_definability}{Lemma}[subsection]
\begin{recursive_definability}[Definability of Recursive Functions]
Every recursive function is arithmetical.
\end{recursive_definability}

It is a well-known fact that there exist one-to-one correspondences between the set of natural numbers and the set of ordered pairs of natural numbers and, indeed, that such a correspondence can be set up in an effective manner \cite[p43]{computability_1}.

\newtheorem{pair_function}[recursive_definability]{Lemma}
\begin{pair_function}[Pairing Function Lemma]
There exist recursive functions $\langle x,y\rangle$, $L(z)$, $R(z)$ such that
\begin{eqnarray*}
  \langle L(z),R(z)\rangle &=& z, \\
  L(\langle x,y\rangle) &=& x, \\
  R(\langle x,y\rangle) &=& y.
\end{eqnarray*}
\end{pair_function}

For notational convenience, we denote $(L(z),R(z))$ by $\bar{z}$. The pairing function can be extended to $n$-tuples by setting
\begin{equation*}
\langle x_1,x_2,\ldots,x_n\rangle = \langle x_1,\langle x_2,\ldots,x_n\rangle\rangle
\end{equation*}
and
\begin{equation*}
\overline{\langle x_1,\ldots.x_n\rangle} = (x_1,\overline{\langle x_2,\ldots,x_n\rangle}).
\end{equation*}

To code a finite sequence $(a_0,\ldots,a_r)$ of natural numbers, we can use two suitably chosen natural numbers $s$ and $t$ as codes, which is the following lemma\cite[p177]{logic_1}.

\newtheorem{beta_function}[recursive_definability]{Lemma}
\begin{beta_function}[$\beta$-Function Lemma]
  There is a recursive function $\beta:N^3\rightarrow N$ satisfying for every sequence $(a_0,\ldots,a_r)$ over $N$ there exist $s,t\in N$ such that for all $i\leq r$, $\beta(s,t,i)=a_i$.
\end{beta_function}

Now, we define $(x)_i$ to be
\begin{equation*}
(x)_i = \beta(L(x),R(x),i).
\end{equation*}
Then, for any finite sequence $(a_0,\ldots,a_r)$ of natural numbers, we can use one natural number $c$ as the code such that $(c)_i = a_i$ for each $i\leq r$. Obviously, $(x)_i$ is a recursive function, which is arithmetical.

\section{Hoare's logic based on $PA$}

The logical language $L'$ for Hoare's logic with respect to Peano's arithmetic consists of the language $L$ of arithmetic, in which assertions describing a program's behavior are specified and the expressions forming the right-hand sides of assignment statements and quantifier-free boolean expressions of conditionals and iterative statements are specified, plus program constructs:
\begin{itemize}
  \item constant symbols: $\textbf{0}$, $\textbf{1}$ ;
  \item function symbols: $+$, $\cdot$ ;
  \item binary predicate symbol: $<$ ;
  \item variables for numbers: $x_0$, $x_1$, $x_2$, $\ldots$ ;
  \item logical connectives and quantifiers: $\neg$, $\rightarrow$, $\forall$ ;
  \item program constructs: $:=$, $;$, $if$, $then$, $else$, $fi$, $while$, $do$, $od$.
\end{itemize}
An expression $e$ is defined as follows:
\begin{equation*}
e ::= \textbf{0} \mid \textbf{1} \mid x \mid e_1+e_2 \mid e_1\cdot e_2;
\end{equation*}
and a boolean expression $b$ is defined as follows:
\begin{equation*}
b ::= e_1 < e_2 \mid \neg b_1 \mid b_1\rightarrow b_2.
\end{equation*}
A program $S$ is defined as follows:
\begin{equation*}
S ::= x := e \mid S_1;S_2 \mid if\ b\ then\ S_1\ else\ S_2\ fi \mid while\ b\ do\ S_0\ od.
\end{equation*}
An assertion $p$ is defined as follows:
\begin{equation*}
p ::= b \mid \neg p_1 \mid p_1\rightarrow p_2 \mid \forall x p_1(x).
\end{equation*}
Then, an asserted program $\varphi$ is a Hoare's triple of the following form:
\begin{equation*}
\varphi ::= \{p_1\} S \{p_2\}.
\end{equation*}

In the present context, we obviously prefer the more suggestive term specification to theory.

Hoare's logic with respect to the specification $PA$, has the following axioms and proof rules for manipulating asserted programs: Let $S,S_0,S_1,S_2\in WP$; $p,q,p_1,q_1,r\in L$; $b\in L$, a quantifier-free formula.

(1) Assignment axiom scheme: for $e$ an expression and $x$ a variable of $L'$, the asserted program
\begin{displaymath}
\begin{array}{l}
\{p(e/x)\}x:=e\{p\}
\end{array}
\end{displaymath}
is an axiom, where $p(e/x)$ stands for the result of substituting $e$ for free occurrences of $x$ in $p$.

(2) Composition rule:
\begin{displaymath}
\begin{array}{c}
  \{p\}S_1\{r\},\{r\}S_2\{q\} \\
  \hline
  \{p\}S_1;S_2\{q\}
\end{array}
\end{displaymath}

(3) Conditional rule:
\begin{displaymath}
\begin{array}{c}
  \{p\wedge b\}S_1\{q\},\{p\wedge \neg b\}S_2\{q\} \\
  \hline
  \{p\}if\ b\ then\ S_1\ else\ S_2\ fi\{q\}
\end{array}
\end{displaymath}

(4) Iteration rule:
\begin{displaymath}
\begin{array}{c}
  \{p\wedge b\}S_0\{p\} \\
  \hline
  \{p\}while\ b\ do\ S_0\ od\{p\wedge \neg b\}
\end{array}
\end{displaymath}

(5) Consequence rule:
\begin{displaymath}
\begin{array}{c}
  p\rightarrow p_1,\{p_1\}S\{q_1\},q_1\rightarrow q \\
  \hline
  \{p\}S\{q\}
\end{array}
\end{displaymath}

And, in connection with (5),

(6) Specification axiom: Each theorem of $PA$ is an axiom.

Note that the purpose of the specification is is to generate the assertions about the data types necessary to govern Consequence rule. The set of asserted programs derivable from these axioms by the proof rules we denote $HL(PA)$ and we write $HL(PA)\vdash \{p\}S\{q\}$ in place of $\{p\}S\{q\}\in HL(PA)$.

The semantics for Hoare's logic is a Kripke's possible world semantics, where the possible worlds are stores, and $S$ are interpreted as modalities.

Let $\textbf{M}=(M,I)$ be a model of $PA$ and $W$ the set of all the assignments(stores), functions from variables to $M$.

A model $\textbf{M}'$ for Hoare's logic based on $PA$ is the triple $(W,\{ R_S: S\in WP \},\textbf{M})$ where $R_S$ is a binary relation on $W$ such that for any $w,w'\in W$,
\begin{itemize}
  \item $(w,w')\in R_{x:=e}$ iff $w'=w(e^{I,w}/x)$, where $e^{I,w}$is defined in the following;
  \item $(w,w')\in R_{S_1;S_2}$ iff $(w,w')\in R_{S_1}\circ R_{S_2}$, where $(z,z')\in R_1\circ R_2$ iff there is a $z''$ such that $(z,z'')\in R_1$ and $(z'',z')\in R_2$;
  \item $(w,w')\in R_{if\ b\ then\ S_1\ else\ S_2\ fi}$ iff either $\textbf{M},w\models b$ and $(w,w')\in R_{S_1}$ or $\textbf{M},w\not\models b$ and $(w,w')\in R_{S_2}$;
  \item $(w,w')\in R_{while\ b\ do\ S_{0}\ od}$ iff there exist natural number $i$ and $w_0=w,w_1,\ldots,w_i=w'$ such that for each $0 \leq j < i$, $\textbf{M},w_j\models b$, $(w_j,w_{j+1})\in R_{S_{0}}$, and $\textbf{M},w'\not\models b$.
\end{itemize}

Given a possible world $w$ and an expression $e$, the interpretation $e^{I,w}$ of $e$ in $w$ is defined as follows:
\begin{displaymath}
  e^{I,w} = \left\{
              \begin{array}{ll}
                \textbf{0}^{\textbf{M}} & \hbox{if $e=\textbf{0}$} \\
                \textbf{1}^{\textbf{M}} & \hbox{if $e=\textbf{1}$} \\
                w(x) & \hbox{if $e=x$} \\
                e_1^{I,w}+^{\textbf{M}}e_2^{I,w} & \hbox{if $e=e_1+e_2$} \\
                e_1^{I,w}\cdot^{\textbf{M}}e_2^{I,w} & \hbox{if $e=e_1\cdot e_2$}.
              \end{array}
            \right.
\end{displaymath}

Given a possible world $w$ and a boolean expression $b$, $b$ is satisfied in $w$, denoted by $\textbf{M},w\models b$, if
\begin{displaymath}
\left\{
  \begin{array}{ll}
    e_1^{I,w} <^{\textbf{M}} e_2^{I,w} & \hbox{if $b = e_1 < e_2$} \\
    \textbf{M},w \not\models b_1 & \hbox{if $b = \neg b_1$} \\
    \textbf{M},w \models b_1 \Rightarrow \textbf{M},w \models b_2, & \hbox{if $b = b_1\rightarrow b_2$},
  \end{array}
\right.
\end{displaymath}
where instead of using $\Rightarrow$ as the logical implication in Hoare's logic, we use $\rightarrow$ \footnote{In syntax, we use $\neg, \wedge, \rightarrow, \forall, \exists$ to denote the logical connectives and quantifiers; and in semantics we use $\sim, \&, \Rightarrow, \textbf{A}, \textbf{E}$ to denote the corresponding connectives and quantifiers.}.

Given a possible world $w$ and an assertion $p$, we say that $p$ is satisfied in $w$, denoted by $\textbf{M},w \models p$, if
\begin{displaymath}
\left\{
  \begin{array}{ll}
    \textbf{M},w \models b & \hbox{if $p = b$} \\
    \textbf{M},w \not\models p_1 & \hbox{if $p = \neg p_1$} \\
    \textbf{M},w \models p_1 \Rightarrow \textbf{M},w \models p_2, & \hbox{if $p = p_1\rightarrow p_2$} \\
    \textbf{A} a\in M (\textbf{M},w(a/x) \models p_1(x)) & \hbox{if $p = \forall x p_1(x)$}.
  \end{array}
\right.
\end{displaymath}

Given a possible world $w$, an asserted program $\{p\} S \{ q \}$ is satisfied at $w$, denoted by $\textbf{M},w\models \{p\}S\{q\}$, if
\begin{equation*}
  \textbf{M},w\models p \Rightarrow Aw'\big((w,w')\in R_S \Rightarrow \textbf{M},w'\models q\big).
\end{equation*}
Then, $\textbf{M}\models \{p\}S\{q\}$ iff for any possible world $w$, $\textbf{M},w\models \{p\}S\{q\}$. $PA\models \{p\}S\{q\}$ iff for any model $\textbf{M}$ of $PA$, $\textbf{M} \models \{p\}S\{q\}$. We also denote $PA\models \{p\}S\{q\}$ by $HL(PA)\models \{p\}S\{q\}$.

\section{Proof of the Separation Theorem}

\subsection{Arithmetical definability of $\textbf{N}$-computable functions}

Let $\textbf{N}'=(W,\{ R_S: S\in WP \},\textbf{N})$ be the model of Hoare's logic based on $PA$, where $R_S$ is the accessibility relation for $S\in WP$ over the standard structure $\textbf{N}$, and $\vec{x}$ the list of variables of length $n$ occurring in while-program $S$ in the following. $R_S$ induces a vectorial function $f_S:N^n\rightarrow N^n$ such that for any $\vec{a},\vec{c}\in N^n$,
\begin{equation*}
f_S(\vec{a}) = \vec{c}
\end{equation*}
if and only if there exist $w,w'\in W$ such that $w(\vec{x})=\vec{a}$, $w'(\vec{x})=\vec{c}$ and $(w,w')\in R_S$. Let $f_S = (f_S^{(1)},\ldots,f_S^{(i)},\ldots,f_S^{(n)})$. Then the $i$th component $f_S^{(i)}$ of $f_S$ is a function from $N^n$ to $N$. We call the function $f_S^{(i)}$ is $\textbf{N}$-computable.

\newtheorem{function_well_defined}{Lemma}[subsection]
\begin{function_well_defined}
$f_S$ is well-defined. That is, for any $w_1,w_1',w_2,w_2' \in W$ with $(w_1,w_1'),(w_2,w_2') \in R_S$, if $w_1(\vec{x})=w_2(\vec{x})$, then $w_1'(\vec{x}) = w_2'(\vec{x})$.
\end{function_well_defined}
\begin{proof}
By induction on the structure of $S$.
\end{proof}

Next, we study how to arithmetically define $f_S$. Construct the formula $\alpha_S$ inductively in $L$.

Assignment: $S \equiv x_i := e$
\begin{equation*}
\alpha_S(\vec{x},\vec{y}) \equiv y_i=e(\vec{x}) \wedge \bigwedge_{j \neq i} y_j=x_j;
\end{equation*}

Composition: $S \equiv S_1;S_2$
\begin{equation*}
\alpha_S(\vec{x},\vec{y}) \equiv \exists \vec{z} (\alpha_{S_1}(\vec{x},\vec{z}/\vec{y})\wedge \alpha_{S_2}(\vec{z}/\vec{x},\vec{y}));
\end{equation*}

Conditional: $S \equiv if\ b\ then\ S_1\ else\ S_2\ fi$
\begin{equation*}
\alpha_S(\vec{x},\vec{y}) \equiv (b(\vec{x})\wedge \alpha_{S_1}(\vec{x},\vec{y})) \vee (\neg b(\vec{x}) \wedge \alpha_{S_2}(\vec{x},\vec{y}));
\end{equation*}

Iteration: $S \equiv while\ b\ do\ S_0\ od$. First, define
\begin{eqnarray*}
  A_S(i,w,\vec{x},\vec{y}) &\equiv& \vec{x}=\overline{(w)_{\textbf{0}}} \wedge \forall j<i (b(\overline{(w)_j}/\vec{x}) \\
                         &&       \wedge \: \alpha_{S_0}(\overline{(w)_j}/\vec{x},\overline{(w)_{j+\textbf{1}}}/\vec{y})) \wedge \vec{y} = \overline{(w)_i}
\end{eqnarray*}
then, set
\begin{equation*}
\alpha_S^*(i,\vec{x},\vec{y}) \equiv \exists w A_S(i,w,\vec{x},\vec{y})
\end{equation*}
and so define
\begin{equation*}
\alpha_S(\vec{x},\vec{y}) \equiv \exists i \alpha_S^*(i,\vec{x},\vec{y}) \wedge \neg b(\vec{y}/\vec{x}).
\end{equation*}

\newtheorem{function_definability}[function_well_defined]{Lemma}
\begin{function_definability}\label{function_definability_label}
$f_S$ is arithmetically definable by $\alpha_S$.
\end{function_definability}
\begin{proof}
It is equivalent to prove that for any $\vec{a},\vec{c}\in N^n$, $f_{S}(\vec{a})=\vec{c}$ $\Leftrightarrow$ $\alpha_S(\vec{\textbf{a}},\vec{\textbf{c}})$ is correct.

The argument is an induction on $S$ for which the basis is the assignment statement.

Assignment:$S\equiv x_i:=e$. $f_S(\vec{a})=\vec{c}$ $\Leftrightarrow$ $(v(\vec{a}/\vec{x}),v(\vec{c}/\vec{x}))\in R_S$ $\Leftrightarrow$ $c_i=e^{I,v(\vec{a}/\vec{x})}$ and $c_j = a_j$ for $j\neq i$ $\Leftrightarrow$ $\mathbf{c_i} = e(\textbf{a})\wedge \bigwedge_{j\neq i}\mathbf{c_j} = \mathbf{a_j}$ is correct $\Leftrightarrow$ $\alpha_S(\vec{\textbf{a}},\vec{\textbf{c}})$ is correct.

Composition:$S\equiv S_1;S_2$. $f_S(\vec{a})=\vec{c}$ $\Leftrightarrow$ $(v(\vec{a}/\vec{x}),v(\vec{c}/\vec{x}))\in R_S$ $\Leftrightarrow$ ($\Rightarrow$.There exists a vector $\vec{d}$ of natural numbers such that)$(v(\vec{a}/\vec{x}),v(\vec{d}/\vec{x}))\in R_{S_1}$ and $(v(\vec{d}/\vec{x}),v(\vec{c}/\vec{x}))\in R_{S_2}$ $\Leftrightarrow$ $f_{S_1}(\vec{a})=\vec{d}$ and $f_{S_2}(\vec{d})=\vec{c}$ $\Leftrightarrow$ ($\Leftarrow$.There exists a vector $\vec{d}$ of natural numbers such that)$\alpha_{S_1}(\vec{\textbf{a}},\vec{\textbf{d}})$ is correct and $\alpha_{S_2}(\vec{\textbf{d}},\vec{\textbf{c}})$ is correct $\Leftrightarrow$ $\exists \vec{z}(\alpha_{S_1}(\vec{\textbf{a}},\vec{z}/\vec{y})\wedge \alpha_{S_2}(\vec{z}/\vec{x},\vec{\textbf{c}}))$ is correct $\Leftrightarrow$ $\alpha_S(\vec{\textbf{a}},\vec{\textbf{c}})$ is correct.

Conditional:$S\equiv if\ b\ then\ S_1\ else\ S_2\ fi$. $f_S(\vec{a})=\vec{c}$ $\Leftrightarrow$ $(v(\vec{a}/\vec{x}),v(\vec{c}/\vec{x}))\in R_S$ $\Leftrightarrow$ $\textbf{N},v(\vec{a}/\vec{x})\models b$ and $(v(\vec{a}/\vec{x}),v(\vec{c}/\vec{x}))\in R_{S_1}$ or $\textbf{N},v(\vec{a}/\vec{x})\not\models b$ and $(v(\vec{a}/\vec{x}),v(\vec{c}/\vec{x}))\in R_{S_2}$ $\Leftrightarrow$ $\textbf{N}\models b(\vec{\textbf{a}})$ and $f_{S_1}(\vec{a})=\vec{c}$ or $\textbf{N}\models \neg b(\vec{\textbf{a}})$ and $f_{S_2}(\vec{a})=\vec{c}$ $\Leftrightarrow$ $b(\vec{\textbf{a}})$ is correct and $\alpha_{S_1}(\vec{\textbf{a}},\vec{\textbf{c}})$ is correct or $\neg b(\vec{\textbf{a}})$ is correct and $\alpha_{S_2}(\vec{\textbf{a}},\vec{\textbf{c}})$ is correct $\Leftrightarrow$ $\big(b(\vec{\textbf{a}})\wedge \alpha_{S_1}(\vec{\textbf{a}},\vec{\textbf{c}})\big)\vee \big(\neg b(\vec{\textbf{a}})\wedge \alpha_{S_2}(\vec{\textbf{a}},\vec{\textbf{c}})\big)$ is correct $\Leftrightarrow$ $\alpha_S(\vec{\textbf{a}},\vec{\textbf{c}})$ is correct.

Iteration:$S\equiv while\ b\ do\ S_0\ od$. $f_S(\vec{a})=\vec{c}$ $\Leftrightarrow$ $(v(\vec{a}/\vec{x}),v(\vec{c}/\vec{x}))\in R_S$ $\Leftrightarrow$ there exist natural number $i$ and vectors of natural numbers $\vec{a_0}$,$\ldots$,$\vec{a_i}$ such that $\vec{a}=\vec{a_0}$, for any $j$ less than $i$, $\textbf{N},v(\vec{a_j}/\vec{x})\models b$ and $(v(\vec{a_j}/\vec{x}),v(\vec{a_{j+1}}/\vec{x}))\in R_{S_0}$, $\vec{c}=\vec{a_i}$, $\textbf{N},v(\vec{c}/\vec{x})\not\models b$ $\Leftrightarrow$ there exist natural number $i$ and vectors of natural numbers $\vec{a_0}$,$\ldots$,$\vec{a_i}$ such that $\vec{a}=\vec{a_0}$, for any $j$ less than $i$, $\textbf{N}\models b(\vec{\mathbf{a_j}})$ and $f_{S_0}(\vec{a_j})=\vec{a_{j+1}}$, $\vec{c}=\vec{a_i}$, $\textbf{N}\models \neg b(\vec{\mathbf{c}})$ $\Leftrightarrow$ there exist natural number $i$ and vectors of natural numbers $\vec{a_0}$,$\ldots$,$\vec{a_i}$ such that $\vec{a}=\vec{a_0}$, for any $j$ less than $i$, $b(\vec{\mathbf{a_j}})$ is correct and $\alpha_{S_0}(\vec{\mathbf{a_j}},\vec{\mathbf{a_{j+1}}})$ is correct, $\vec{c}=\vec{a_i}$, $\neg b(\vec{\mathbf{c}})$ is correct $\Leftrightarrow$ there exist natural number $i$ and vectors of natural numbers $\vec{a_0}$,$\ldots$,$\vec{a_i}$ such that $\vec{\mathbf{a}}=\vec{\mathbf{a_0}}\wedge \forall j<\textbf{i}\Big(b(\vec{\mathbf{a_j}})\wedge \alpha_{S_0}\big(\vec{\mathbf{a_j}}, \vec{\mathbf{a_{j+1}}}\big)\Big)\wedge \vec{\mathbf{c}}=\vec{\mathbf{a_i}} \wedge \neg b(\vec{\mathbf{c}})$ is correct $\Leftrightarrow$ there exist natural numbers $i$, $d_0$,$\ldots$,$d_i$ such that $\vec{\mathbf{a}}=\overline{\mathbf{d_0}}\wedge \forall j<\textbf{i}\Big(b(\overline{\mathbf{d_j}})\wedge \alpha_{S_0}\big(\overline{\mathbf{d_j}}, \overline{\mathbf{d_{j+1}}}\big)\Big)\wedge \vec{\mathbf{c}}=\overline{\mathbf{d_i}} \wedge \neg b(\vec{\mathbf{c}})$ is correct $\Leftrightarrow$ there exist natural numbers $i$ and $w$ such that $\vec{\mathbf{a}}=\overline{\mathbf{(w)_0}}\wedge \forall j<\textbf{i}\Big(b(\overline{\mathbf{(w)_j}})\wedge \alpha_{S_0}\big(\overline{\mathbf{(w)_j}}, \overline{\mathbf{(w)_{j+1}}}\big)\Big)\wedge \vec{\mathbf{c}}=\overline{\mathbf{(w)_i}} \wedge \neg b(\vec{\mathbf{c}})$ is correct $\Leftrightarrow$ there exist natural numbers $i$ and $w$ such that $A_S(\mathbf{i},\mathbf{w},\vec{\mathbf{a}},\vec{\mathbf{c}})\wedge \neg b(\vec{\mathbf{c}})$ is correct $\Leftrightarrow$ $\exists i \exists w B(i,w,\vec{\mathbf{a}},\vec{\mathbf{c}})\wedge \neg b(\vec{\mathbf{c}})$ is correct $\Leftrightarrow$ $\alpha_S(\vec{\mathbf{a}},\vec{\mathbf{c}})$ is correct.

\end{proof}

Define $\alpha_S^{(i)}(\vec{x},y)$ as follows:
\begin{eqnarray*}
   \alpha_S^{(i)}(\vec{x},y) &::=& \exists y_1,\ldots,y_{i-1},y_{i+1},\ldots,y_n \alpha_S(\vec{x}, \\
                                   &&  (y_1,\ldots,y_{i-1},y,y_{i+1},\ldots,y_n)).
\end{eqnarray*}
Then, we can easily achieve that $f_S^{(i)}$ is arithmetically defined by $\alpha_S^{(i)}$ from the above lemma. Actually, these two definabilities are interconvertible.

\subsection{The definition of $SP(p,S)$}

Now, we define strongest postconditions. Given the model $\textbf{N}'=(W,\{ R_S: S\in WP \},\textbf{N})$ of Hoare's logic based on $PA$, $p\in L$ and $S\in WP$, then the strongest postcondition $sp_{\textbf{N}}(p,S)$ of $p$ and $S$ relative to $\textbf{N}$ is the set of possible worlds:
\begin{equation*}
     \{w \in W : \textbf{E}w'\in W(\textbf{N},w'\models p\ \&\ (w',w)\in R_S)\}.
\end{equation*}
From the definitions, it immediately follows that
\newtheorem{strongest_postcondition}{Lemma}[subsection]
\begin{strongest_postcondition}
  $\textbf{N}\models \{p\}S\{q\}\Leftrightarrow sp_{\textbf{N}}(p,S)\subset \{w\in W:\textbf{N},w\models q\}.$
\end{strongest_postcondition}

Like the definition of $SP(p,S)$ in \cite[p271]{bergstra_1}, we can inductively define $SP(p,S)$ over the structure of $S$ as follows:

Assignment: $S \equiv x_i := e$
\begin{equation*}
SP(p,S)(\vec{x}) \equiv \exists \vec{u} \big(p(\vec{u}/\vec{x})\wedge x_i=e(\vec{u}/\vec{x}) \wedge \bigwedge_{j \neq i} x_j=u_j \big);
\end{equation*}

Composition: $S \equiv S_1;S_2$
\begin{equation*}
SP(p,S)(\vec{x}) \equiv SP\big(SP(p,S_1),S_2\big)(\vec{x});
\end{equation*}

Conditional: $S \equiv if\ b\ then\ S_1\ else\ S_2\ fi$
\begin{equation*}
SP(p,S)(\vec{x}) \equiv SP(p\wedge b,S_1)(\vec{x})\vee SP(p\wedge \neg b,S_2)(\vec{x});
\end{equation*}

Iteration: $S \equiv while\ b\ do\ S_0\ od$
\begin{equation*}
SP(p,S)(\vec{x}) \equiv INV(p,b,S_0)(\vec{x})\wedge \neg b(\vec{x}),
\end{equation*}
where $INV(p,b,S_0)$ is the loop invariant formula(refer to Invariant Law in \cite[p277]{bergstra_1}) built up as follows.

First, define
\begin{eqnarray*}
  B(p,b,S_0)(i,w,\vec{x}) &\equiv& p(\overline{(w)_{\textbf{0}}}/\vec{x})\wedge \forall j<i\ SP\big(\vec{x}=\overline{(w)_j} \\
   && \wedge \: b, S_0\big) \big(\overline{(w)_{j+\textbf{1}}}/\vec{x}\big)\wedge \vec{x}=\overline{(w)_i}.
\end{eqnarray*}
Next set
\begin{equation*}
    INV^*(p,b,S_0)(i,\vec{x})\equiv \exists w B(p,b,S_0)(i,w,\vec{x})
\end{equation*}
and so define
\begin{equation*}
    INV(p,b,S_0)(\vec{x})\equiv \exists i INV^*(p,b,S_0)(i,\vec{x}).
\end{equation*}

The definition of $SP(p,S)$ in \cite[p271]{bergstra_1} entails:
\newtheorem{postcondition_definability}[strongest_postcondition]{Lemma}
\begin{postcondition_definability}
$SP(p,S)$ arithmetically defines $sp_{\textbf{N}}(p,S)$.
\end{postcondition_definability}

From the preceding two lemmas, it deduces that
\newtheorem{equivalent_in_N}[strongest_postcondition]{Lemma}
\begin{equivalent_in_N}\label{equivalent_in_N_label}
$\textbf{N}\models \{p\}S\{q\}\Leftrightarrow \textbf{N}\models SP(p,S)\rightarrow q$.
\end{equivalent_in_N}

Lemma \ref{equivalent_in_N_label} is a special case of statement (3) in Theorem \ref{bergstra_theorem}, when restricting the model of $PA$ to $\textbf{N}$.

\subsection{Proof of the Separation Theorem: The equivalent form of $SP(p,S)$ in $PA$}

Until now, we have finished all the preparations for the Separation Theorem. Let's turn to the last part of the proof of the theorem. In the sequel, we argue that $SP(p,S)$ has an equivalent form $\exists \vec{u} ( p(\vec{u}/\vec{x}) \wedge \alpha_S(\vec{u}/\vec{x},\vec{x}/\vec{y}) )$ in $PA$.

\newtheorem{core_of_separation_theorem}{Theorem}[subsection]
\begin{core_of_separation_theorem}
$PA \vdash SP(p,S)(\vec{x}) \leftrightarrow \exists \vec{u} ( p(\vec{u}/\vec{x}) \wedge \alpha_S(\vec{u}/\vec{x},\vec{x}/\vec{y}) )$.
\end{core_of_separation_theorem}
\begin{proof}
By G\"{o}del's Soundness and Completeness Theorem,
\begin{equation*}
PA\vdash SP(p,S)(\vec{x})\leftrightarrow \exists \vec{u}\big(p(\vec{u}/\vec{x})\wedge \alpha_S(\vec{u}/\vec{x},\vec{x}/\vec{y})\big)
\end{equation*}
if and only if
\begin{equation*}
PA\models SP(p,S)(\vec{x})\leftrightarrow \exists \vec{u}\big(p(\vec{u}/\vec{x})\wedge \alpha_S(\vec{u}/\vec{x},\vec{x}/\vec{y})\big).
\end{equation*}
Consequently, it suffices to prove for any $\textbf{M}\models PA$,
\begin{equation*}
\textbf{M}\models SP(p,S)(\vec{x})\leftrightarrow \exists \vec{u}\big(p(\vec{u}/\vec{x})\wedge \alpha_S(\vec{u}/\vec{x},\vec{x}/\vec{y})\big).
\end{equation*}
Fix $\textbf{M}\models PA$. The argument is by induction on the structure of $S$ for which the basis is the assignment.

Assignment:$S\equiv x_i:=e$. According to the definition, $SP(p,S)(\vec{x})$ is equivalent in $\textbf{M}$ to
\begin{equation*}
\exists \vec{u} \big(p(\vec{u}/\vec{x})\wedge x_i=e(\vec{u}/\vec{x}) \wedge \bigwedge_{j \neq i} x_j=u_j \big),
\end{equation*}
by pure logic and the definition of $\alpha_S(\vec{x},\vec{y})$, it is equivalent to
\begin{equation*}
  \exists \vec{u}\big(p(\vec{u}/\vec{x})\wedge \alpha_S(\vec{u}/\vec{x},\vec{x}/\vec{y})\big).
\end{equation*}
Thus,
\begin{equation*}
\textbf{M} \models SP(p,S)(\vec{x})\leftrightarrow  \exists \vec{u}\big(p(\vec{u}/\vec{x})\wedge \alpha_S(\vec{u}/\vec{x},\vec{x}/\vec{y})\big).
\end{equation*}

Composition: $S\equiv S_1;S_2$. The induction hypothesis applied to $S_1$ and $S_2$ yields
\begin{eqnarray*}
  \textbf{M} &\models& SP(p,S_1)(\vec{x})\leftrightarrow \exists \vec{u}\big(p(\vec{u}/\vec{x})\wedge \alpha_{S_1}(\vec{u}/\vec{x},\vec{x}/\vec{y})\big), \\
  \textbf{M} &\models& SP(SP(p,S_1),S_2)(\vec{x}) \\
  &&\leftrightarrow  \exists \vec{u}\big(SP(p,S_1)(\vec{u}/\vec{x})\wedge \alpha_{S_2}(\vec{u}/\vec{x},\vec{x}/\vec{y})\big).
\end{eqnarray*}
Consider this last formula through several transformations: replacing $\vec{u}$ by $\vec{z}$, it is equivalent in $\textbf{M}$ to
\begin{equation*}
\exists \vec{z}\big(SP(p,S_1)(\vec{z}/\vec{x})\wedge \alpha_{S_2}(\vec{z}/\vec{x},\vec{x}/\vec{y})\big).
\end{equation*}
With the substitution of $\exists \vec{u}\big(p(\vec{u}/\vec{x})\wedge \alpha_{S_1}(\vec{u}/\vec{x},\vec{z}/\vec{y})\big)$ for $SP(p,S_1)(\vec{z}/\vec{x})$, it is equivalent to
\begin{equation*}
\exists \vec{z}\Big(\exists \vec{u}\big(p(\vec{u}/\vec{x})\wedge \alpha_{S_1}(\vec{u}/\vec{x},\vec{z}/\vec{y})\big)\wedge \alpha_{S_2}(\vec{z}/\vec{x},\vec{x}/\vec{y})\Big).
\end{equation*}
By pure logic, it is equivalent to
\begin{equation*}
\exists \vec{u}\Big(p(\vec{u}/\vec{x})\wedge \exists \vec{z}\big(\alpha_{S_1}(\vec{u}/\vec{x},\vec{z}/\vec{y})\wedge \alpha_{S_2}(\vec{z}/\vec{x},\vec{x}/\vec{y})\big)\Big).
\end{equation*}
By the definition of $\alpha_S(\vec{x},\vec{y})$, it is equivalent to
\begin{equation*}
\exists \vec{u}\big(p(\vec{u}/\vec{x})\wedge \alpha_S(\vec{u}/\vec{x},\vec{x}/\vec{y})\big).
\end{equation*}
Thus, by the definition, $SP(p,S)(\vec{x})$ is equivalent in $\textbf{M}$ to
\begin{equation*}
SP(SP(p,S_1),S_2)(\vec{x}),
\end{equation*}
By the above argument, it is equivalent to
\begin{equation*}
\exists \vec{u}\big(p(\vec{u}/\vec{x})\wedge \alpha_S(\vec{u}/\vec{x},\vec{x}/\vec{y})\big).
\end{equation*}
Therefore,
\begin{equation*}
\textbf{M} \models SP(p,S)(\vec{x})\leftrightarrow  \exists \vec{u}\big(p(\vec{u}/\vec{x})\wedge \alpha_S(\vec{u}/\vec{x},\vec{x}/\vec{y})\big).
\end{equation*}

Conditional: $S\equiv if\ b\ then\ S_1\ else\ S_2\ fi$. By the induction hypothesis applied to $S_1$, $S_2$, it yields
\begin{eqnarray*}
  \textbf{M} &\models& SP(p\wedge b,S_1)(\vec{x})\\
  &&\leftrightarrow \exists \vec{u}\big(p(\vec{u}/\vec{x})\wedge b(\vec{u}/\vec{x})\wedge \alpha_{S_1}(\vec{u}/\vec{x},\vec{x}/\vec{y})\big), \\
  \textbf{M} &\models& SP(p\wedge \neg b,S_2)(\vec{x})\\
  &&\leftrightarrow \exists \vec{u}\big(p(\vec{u}/\vec{x})\wedge \neg b(\vec{u}/\vec{x})\wedge \alpha_{S_2}(\vec{u}/\vec{x},\vec{x}/\vec{y})\big).
\end{eqnarray*}
According to the definition, $SP(p,S)(\vec{x})$ is equivalent in $\textbf{M}$ to
\begin{equation*}
SP(p\wedge b,S_1)(\vec{x})\vee SP(p\wedge \neg b,S_2)(\vec{x})
\end{equation*}
Consider this last formula through several transformations: by substitutions, it is equivalent in $\textbf{M}$ to
\begin{eqnarray*}
  && \exists \vec{u}\big(p(\vec{u}/\vec{x})\wedge b(\vec{u}/\vec{x})\wedge \alpha_{S_1}(\vec{u}/\vec{x},\vec{x}/\vec{y})\big) \\
  && \vee \: \exists \vec{u}\big(p(\vec{u}/\vec{x})\wedge \neg b(\vec{u}/\vec{x})\wedge \alpha_{S_2}(\vec{u}/\vec{x},\vec{x}/\vec{y})\big).
\end{eqnarray*}
By pure logic, it is equivalent to
\begin{eqnarray*}
  && \exists \vec{u}\bigg(p(\vec{u}/\vec{x})\wedge \Big(\big(b(\vec{u}/\vec{x})\wedge \alpha_{S_1}(\vec{u}/\vec{x},\vec{x}/\vec{y})\big) \\
  &&\vee \: \big(\neg b(\vec{u}/\vec{x})\wedge \alpha_{S_2}(\vec{u}/\vec{x},\vec{x}/\vec{y})\big)\Big)\bigg).
\end{eqnarray*}
By the definition of $\alpha_S(\vec{x},\vec{y})$, it is equivalent to
\begin{equation*}
\exists \vec{u}\big(p(\vec{u}/\vec{x})\wedge \alpha_S(\vec{u}/\vec{x},\vec{x}/\vec{y})\big).
\end{equation*}
So, we obtain
\begin{equation*}
\textbf{M} \models SP(p,S)(\vec{x})\leftrightarrow  \exists \vec{u}\big(p(\vec{u}/\vec{x})\wedge \alpha_S(\vec{u}/\vec{x},\vec{x}/\vec{y})\big).
\end{equation*}

Iteration: $S\equiv while\ b\ do\ S_0\ od$. Applying the induction hypothesis to $S_0$ yields
\begin{eqnarray*}
  \textbf{M} &\models& SP\big(\vec{x}=\overline{(w)_j}\wedge b,S_0\big)(\vec{x}) \\
  && \leftrightarrow \exists \vec{u}\big(\vec{u}=\overline{(w)_j}\wedge b(\vec{u}/\vec{x})\wedge \alpha_{S_0}(\vec{u}/\vec{x},\vec{x}/\vec{y})\big).
\end{eqnarray*}
By the generalized coding function assumed by Bergstra, et.al\cite{bergstra_1}, the last formula in the above theorem is equivalent in $\textbf{M}$ to
\begin{equation*}
b\big(\overline{(w)_j}/\vec{x}\big)\wedge \alpha_{S_0}\big(\overline{(w)_j}/\vec{x},\vec{x}/\vec{y}\big).
\end{equation*}
Then,
\begin{eqnarray*}
  \textbf{M} &\models& SP\big(\vec{x}=\overline{(w)_j}\wedge b,S_0\big)(\vec{x}) \\
  && \leftrightarrow b\big(\overline{(w)_j}/\vec{x}\big)\wedge \alpha_{S_0}\big(\overline{(w)_j}/\vec{x},\vec{x}/\vec{y}\big).
\end{eqnarray*}
Substituting $\overline{(w)_{j+\textbf{1}}}$ for each free occurrence of $\vec{x}$ in the above formula gives
\begin{eqnarray*}
  \textbf{M} &\models& SP\big(\vec{x}=\overline{(w)_j}\wedge b,S_0\big)\big(\overline{(w)_{j+\textbf{1}}}/\vec{x}\big) \\
  && \leftrightarrow b\big(\overline{(w)_j}/\vec{x}\big)\wedge \alpha_{S_0}\big(\overline{(w)_j}/\vec{x},\overline{(w)_{j+\textbf{1}}}/\vec{y}\big).
\end{eqnarray*}
From the definition, it follows directly that $B(p,b,S_0)(i,w,\vec{x})$ is equivalent in $\textbf{M}$ to
\begin{eqnarray*}
  && p(\overline{(w)_{\textbf{0}}}/\vec{x}) \wedge \forall j<i SP\big(\vec{x}=\overline{(w)_j}\wedge b,S_0\big)\big(\overline{(w)_{j+\textbf{1}}}/\vec{x}\big) \\
  && \wedge \: \vec{x}=\overline{(w)_i}.
\end{eqnarray*}
By the above formula, it is equivalent to
\begin{eqnarray*}
  && p(\overline{(w)_{\textbf{0}}}/\vec{x}) \wedge \forall j<i \Big(b\big(\overline{(w)_j}/\vec{x}\big) \\
  && \wedge \: \alpha_{S_0}\big(\overline{(w)_j}/\vec{x},\overline{(w)_{j+\textbf{1}}}/\vec{y}\big)\Big)\wedge \vec{x}=\overline{(w)_i}.
\end{eqnarray*}
According to the definition of $A_S(i,w,\vec{x},\vec{y})$, it is equivalent to
\begin{equation*}
    p(\overline{(w)_{\textbf{0}}}/\vec{x})\wedge A_S(i,w,\overline{(w)_{\textbf{0}}}/\vec{x},\vec{x}/\vec{y}).
\end{equation*}
With the generalized coding function, it is equivalent in $\textbf{M}$ to
\begin{equation*}
\exists \vec{u}\big(p(\vec{u}/\vec{x})\wedge A_S(i,w,\vec{u}/\vec{x},\vec{x}/\vec{y})\big).
\end{equation*}
Therefore,
\begin{eqnarray*}
  \textbf{M} &\models& B(p,b,S_0)(i,w,\vec{x})  \\
   &&  \leftrightarrow \exists \vec{u}\big(p(\vec{u}/\vec{x})\wedge A_S(i,w,\vec{u}/\vec{x},\vec{x}/\vec{y})\big).
\end{eqnarray*}
By the definition, $SP(p,S)(\vec{x})$ is equivalent in $\textbf{M}$ to
\begin{equation*}
INV(p,b,S_0)(\vec{x})\wedge \neg b(\vec{x}).
\end{equation*}
Now, consider the last formula through several transformations: it is equivalent in $\textbf{M}$ to
\begin{equation*}
\exists i \exists w B(p,b,S_0)(i,w,\vec{x}) \wedge \neg b(\vec{x}).
\end{equation*}
By the above formula, it is equivalent to
\begin{equation*}
\exists i \exists w \exists \vec{u}\big(p(\vec{u}/\vec{x})\wedge A_S(i,w,\vec{u}/\vec{x},\vec{x}/\vec{y})\big) \wedge \neg b(\vec{x}).
\end{equation*}
By pure logic, it is equivalent to
\begin{equation*}
\exists \vec{u}\big(p(\vec{u}/\vec{x})\wedge \exists i \exists w A_S(i,w,\vec{u}/\vec{x},\vec{x}/\vec{y}) \wedge \neg b(\vec{x})\big).
\end{equation*}
And, clearly, by the definition of $\alpha_S(\vec{x},\vec{y})$, this last formula is equivalent in $\textbf{M}$ to
\begin{equation*}
\exists \vec{u}\big(p(\vec{u}/\vec{x})\wedge \alpha_S(\vec{u}/\vec{x},\vec{x}/\vec{y})\big).
\end{equation*}
Thus,
\begin{equation*}
\textbf{M}\models SP(p,S)(\vec{x})\leftrightarrow \exists \vec{u}\big(p(\vec{u}/\vec{x})\wedge \alpha_S(\vec{u}/\vec{x},\vec{x}/\vec{y})\big).
\end{equation*}
And we are done.
\end{proof}

\section{Discussion and Future Work}

The corollary \ref{from_hl_to_pa_label} looks on $\alpha_S$ as the logical characterization of while-program $S$ in $PA$. In the sequel, we shall extend the computability induced by $S$ from the standard structure to any nonstandard model of $PA$ through $\alpha_S$. Lemma \ref{function_definability_label} presents that $\alpha_S$ arithmetically defines $f_{S}^{\textbf{N}}$, i.e. for any vectors of natural numbers $\vec{a}$,$\vec{c}$, $f_S^{\textbf{N}}(\vec{a})=\vec{c}$ iff $\textbf{N}\models \alpha_S(\vec{a},\vec{c})$. Then for any nonstandard model $\textbf{M}=(M,I)$ of $PA$, we define $f^{\textbf{M}}_{S}$ such that for any elements $\vec{a},\vec{c}\in M^n$, $f^{\textbf{M}}_S(\vec{a})=\vec{c}$ iff $\textbf{M} \models \alpha_S(\vec{a},\vec{c})$. Thus, $f^{\textbf{M}}_{S}$ is an extension of $f^{\textbf{N}}_{S}$ with the uniform $\Sigma_1$-definability in $L$. It remains to verify the validity of this underlying semantics(type-1 semantics).

Consider the while-program $S$ as follows:
\begin{equation*}
y:=\textbf{0};while\ y<x\ do\ y:=y+\textbf{1}\ od,
\end{equation*}
where the program variable $x$ is the only input and output variable. Then, by the generalized coding function, $f_S^{\textbf{N}}=id^{\textbf{N}}$ and $\alpha_S(x,y)$ is logically equivalent to $y=x$. By $id^{\textbf{N}}$ we mean the identity function defined in $N$, i.e. $dom(id^{\textbf{N}})=N$. Let $\textbf{M}=(M,I)$ be any nonstandard model of $PA$. Then, $\textbf{M}$-computable function $f^{\textbf{M}}_{S}$ defined in $\textbf{M}$ by $\alpha_S$, in which the generalized coding function codes the $\textbf{M}$-finite sequence of pairs $(0,x)$, $(1,x)$ $\ldots$ $(x,x)$, is $id^{\textbf{M}}$($dom(id^{\textbf{M}})=M$). Therefore, the type-1 semantics extends the number of halting steps from $\textbf{N}$-finiteness to $\textbf{M}$-finiteness, which makes $L$ expressive relative to $\textbf{M}$.

On the other hand, up to now, we have known that the number of iterative steps of a loop is defined in the standard part of $\textbf{M}$, as treated in this paper. In this semantics(type-2 semantics), the $\textbf{M}$-computable function induced by $S$ on $\textbf{M}$ is defined in $N^{\textbf{M}}$, i.e. $id^{\textbf{N}^{\textbf{M}}}$. By the Overspill principle\cite[p309]{c. and l.}, $N^{\textbf{M}}$ is not arithmetically defined in $\textbf{M}$. Hence, $id^{\textbf{N}^{\textbf{M}}}$ is not arithmetically defined in $\textbf{M}$.

To make the type-2 semantics defined logically, introduce a new predicate symbol to the language of arithmetic, which defines $N^{\textbf{M}}$ in $\textbf{M}$, and restrict its occurrences in formulas, say in a guarded fashion. The paper \cite{cungen_1} is an attempt for the type-2 semantics.

\section*{Acknowledgement}
I gratefully thank Prof. Yuefei Sui for many long and interesting discussions about Hoare's logic, and for review of this paper. I also thank Prof. Wenhui Zhang for comments on earlier versions of this paper.





\bibliographystyle{model1-num-names}
\bibliography{A New Perspective for Hoare's Logic and Peano's Arithmetic}

\begin{thebibliography}{99}

\bibitem{apt_1}
K.~R.~Apt,
\newblock Ten year's of Hoare's logic: a survey,
\newblock ACM Trans. Programming Languages and Systems 3(4) (1981) 431-483.
\bibitem{hoare_1}
C.~A.~R.~Hoare,
\newblock An axiomatic basis for computer programming,
\newblock Comm. ACM 12 (1969) 576-580.
\bibitem{cook_1}
S.~A.~Cook,
\newblock Soundness and completeness of an axiom system for program verification,
\newblock SIAM J. Comput. 7 (1978) 70-90.
\bibitem{lipton_1}
R.~J.~Lipton,
\newblock A necessary and sufficient condition for the existence of Hoare Logics,
\newblock In Proc. 18th IEEE Symp. Foundations of Computer Science (1977) 1-6.
\bibitem{wand_1}
M.~Wand,
\newblock A new incompleteness result for Hoare's system,
\newblock J. ACM 25(1) (1978) 168-175.
\bibitem{clarke_1}
E.~M.~Clarke~JR.,
\newblock Programming language constructs for which it is impossible to obtain good Hoare axiom systems,
\newblock J. ACM  26(1) (1979) 129-147.
\bibitem{reynolds_1}
J.~C.~Reynolds,
\newblock Separation logic: a logic for shared mutable data structures,
\newblock LICS (2002) 55-74.
\bibitem{o'hearn_1}
P.~W.~O'Hearn, H.~Yang, J.~C.~Reynolds,
\newblock Separation and information hiding,
\newblock in Proceedings of POPL (2004) 268-280.
\bibitem{parkinson_1}
M.~Parkinson, G.~Bierman,
\newblock Separation logic and abstraction,
\newblock in Proceedings of POPL (2005) 247-258.
\bibitem{modal_1}
G.~E.~Hughes, M.~J.~Cresswell,
\newblock A new introduction to modal logic,
\newblock Routledge, London, 1996.
\bibitem{bergstra_1}
J.~A.~Bergstra, J.~V.~Tucker,
\newblock Hoare's logic and peano's arithmetic,
\newblock Theoretical Computer Science 22 (1983) 265-284.
\bibitem{c. and l.}
G.~S.~Boolos, J.~P.~Burgess, R.~C.~Jeffrey,
\newblock Computability and logic, fifth edition,
\newblock Cambridge University Press, 2007.
\bibitem{computability_1}
Martin~Davis,
\newblock Computability \& unsovability,
\newblock Courier Dover Publications, New York, 1982.
\bibitem{logic_1}
H.~-D.~Ebbinghaus, J.~Flum, W.~Thomas,
\newblock Mathematical logic, second ed.,
\newblock Springer, New York, 1994.
\bibitem{cungen_1}
C.~Cao, Y.~Sui, Z.~Zhang,
\newblock The $\textbf{M}$-computations induced by accessibility relations in nonstandard models $\textbf{M}$ of Hoare logic,
\newblock Institute of Computing Technology, Chinese Academy of Sciences, 2013.
\bibitem{set_1}
K.~Hrbacek, T.~Jech,
\newblock Introduction to Set Theory, third ed.,
\newblock Marcel Dekker, New York, 1999.
\bibitem{rogers_1}
H.~Rogers,
\newblock Theory of Recursive Functions and Effective Computability,
\newblock The MIT Press, 1987.
\bibitem{barwise}
J.~Barwise,
\newblock Handbook of Mathematical Logic,
\newblock North-Holland, Amsterdam, 1977.

\end{thebibliography}

\end{document}